\documentclass[%
 reprint,
 twocolumn,
superscriptaddress,
nofootinbib,
 showpacs,
 amsmath,amssymb,
 aps,
 prl,
floatfix,
]{revtex4-2}

\usepackage{graphicx}
\graphicspath{{./Fig/}}
\usepackage{amssymb,amsmath,amsthm}
\usepackage{hyperref}
\usepackage[utf8]{inputenc}
\usepackage[english]{babel}

\newmuskip\pFqmuskip

\newcommand*\pFq[6][8]{%
  \begingroup % only local assignments
  \pFqmuskip=#1mu\relax
  \mathchardef\normalcomma=\mathcode`,
  \mathcode`\,=\string"8000
  \begingroup\lccode`\~=`\,
  \lowercase{\endgroup\let~}\pFqcomma
  {}_{#2}F_{#3}{\left(\genfrac..{0pt}{}{#4}{#5};#6\right)}%
  \endgroup
}
\newcommand{\pFqcomma}{{\normalcomma}\mskip\pFqmuskip}%

\begin{document}

\title{Variational Pattern Selection}

\author{Mark Mineev-Weinstein}
\affiliation{New Mexico Consortium, Los Alamos, NM, 87544, USA}
\author{Oleg Alekseev}
\affiliation{Chebyshev Laboratory, Department of Mathematics and Computer Science, Saint-Petersburg State University, 14-th line, 199178, Saint-Petersburg, Russia}

\date{\today}

\begin{abstract}
We address pattern selection problems in non-linear interface dynamics by maximizing the entropy of the most probable (classical) scenario associated with the processes. We applied this variational principle to well-known selection problems in a Hele-Shaw cell: a self-similar finger in a wedge \cite{89} and the stationary Saffman-Taylor finger in a channel \cite{ST58} as a limiting case. The results obtained are excellently consistent with the experiments. We also address the universal fjord opening angle \cite{2OO6}. Surface tension is not needed for the selection, contrary to common belief.
\end{abstract}


\maketitle

Of the many unstable non-linear phenomena, we will address here only those describing front propagation, pattern formation, and self-organization. In addition to being of obvious interests for mathematics, physics, geophysics, chemistry, and biology, they are indispensable in oil/gas recovery, metallurgy, and medicine (malignant growth), to name just a few applications. These processes often exhibit the problem of selecting the most stable member from a family of stationary or self-similar solutions. 

\textit{List of pioneers}.
The first work on pattern selection was on gene propagation (Kolmogorov et al. \cite{K37}). Based on citations, selections of fluid/fluid interfaces named after Taylor (Rayleigh-Taylor \cite{RT} and Saffman-Taylor \cite{ST58} instabilities) attracted most of the attention. Comparable efforts were also made to calculate asymptotic velocities in flame propagation (Landau \cite{Lflame}, Zeldovich \cite{Zflame}, Gelfand \cite{Gflame}), crack propagation (Mott \cite{Mcracks}, Barenblatt \cite{bar}) and dendritic growth (Ivantsov \cite{iv}, Langer \cite{L8O}). These impressive names reflect both the importance and difficulty of selection problems.

\textit{Extremum principle}.
Selection problems are challenging in both physics (identifying the selection mechanism) and mathematics (handling a small singular term). The desire to find a functional, whose extremal describes the selected pattern, is understandable. However, since all of these processes are out of equilibrium, minimizing the thermodynamic potentials here is not helpful. The importance of the extremum principle was clearly indicated by one of the pioneers in dendritic growth selection, J. S. Langer: After noticing ``[t]he big, unsolved part of the problem is how these complex shapes are selected'', he asks: ``\ldots might there be some meaningful and useful variational formulation that describes these processes?'' \cite{L8O}. 

\textit{Minimal dissipation is out of the question}.
There is one such functional applicable out of equilibrium---the principle of minimal dissipation in viscous hydrodynamics, discovered by Helmholtz and Korteveg in the XIX century (see \cite{Lamb}), and rediscovered by Onsager in kinetics of weakly non-equilibrium systems \cite{O}. In fact, the authors of \cite{ST58} tried to apply this functional to select the observed inviscid finger with a relative width of 1/2 in a Hele-Shaw cell. However, it turned out that the complete dissipation (friction between plates and oil) does not depend on the finger shape at all, and thus is of no help. This was one of the motivations to include surface tension as the largest factor neglected in deriving a continuum family of stationary fingers to solve the selection problem. However, inclusion of surface tension is non-trivial because the term with surface tension provides a singular perturbation.

\textit{Selection through surface tension}. This challenge was addressed by using the WKB approximation, originally developed in quantum mechanics \cite{LL-3}, which draws parallels between the surface tension term and the Planck constant in the Schrödinger equation. WKB method helped to answer many questions in quantum mechanics, such as quantum tunneling, Bohr-Sommerfeld quantization, and presenting the action as an adiabatic invariant. 

Kruskal and Segur extended the WKB method with their ``Asymptotics beyond all orders'' theory \cite{KS}, capturing exponentially small terms through analytic continuation. These results culminated in 1986 when several independent groups simultaneously reported in (the same issue of) PRL \cite{PRL-Shraiman,PRL-JSL,PRL-ENS} the successful completion of the Saffman-Taylor finger (STF) selection problem (see also \cite{KKL, NATO-book,Saffman-1986,Saffman-1991}).

\textit{Selection without surface tension}. Later, in 1998 it was shown that the selection of STF does not require surface tension \cite{98}, \cite{Matkowsky}. The result \cite{98} was achieved due to the integrability of Laplacian growth (LG) \cite{OO}. This remarkable property, which allows one to obtain various classes of exact unsteady solutions, is an exception rather than a rule among nonlinear dynamical systems. Integrability, as a new branch of mathematical physics, was born in 1967 after the discovery of striking and unusual (at that time) properties of the Korteveg de Vries equation \cite{ggkm}. This field has been booming since then, with no signs of slowing to date. 

The result \cite{98} was also achieved due to Tikhonov regularization\footnote{There is no reference to Tikhonov in \cite{98}, since the author of \cite{98} was unaware of this regularization in 1998.}, which converted the ill-posed problem into the well-posed one \cite{ti1,ti2,ti3,ti4,ti5}. The approach developed in \cite{98} was later successfully applied to the bubble selection in a channel \cite{TS-1,Robb,TS-2,2O24}.

\textit{Goals of the paper}. Being dissipative, the Hele-Show flow does not possess the action, which minimum provides its dynamics in the form of Euler-Lagrange equations. However, earlier we developed a stochastic theory \cite{O1,O2}, where we obtained dissipative deterministic growth equations by maximizing an entropy of {\it stochastic} processes. In this article, we maximize the entropy of {\it deterministic} processes to address well-known selection problems in a Hele-Shaw cell in a wedge and in a channel geometries. The results we obtained below excellently agree with the selection by surface tension \cite{Brener90,Tu} and with experiments \cite{89,ST58,2OO6}.

\textit{Entropy functional in LG.} Assume $K$ tiny Brownian particles of the area $\hbar$, issued from infinity and land on a boundary of a growing domain $D(t)$ within each time unit \cite{O1,O2}. This process is a bridge (crossover) between $K=1$, describing DLA process \cite{DLA}, and $K\to \infty$, $\hbar \to 0$, which corresponds to a deterministic Laplacian growth (LG) described by the bilinear equation \eqref{lg-equation} below. The correspondence with the classical limit implies the condition
\begin{equation}\label{sc-limit}
    K \hbar  = Q \delta t. 
\end{equation}
This is the total area of particles attached to the boundary of $D(t)$ per time unit $\delta t$, that is, the area of a layer grown during $\delta t$. Hence, $Q$ is a strength of a fluid source in deterministic Laplacian growth.

By dividing the boundary of $D(t)$ into $N\gg1$ tiny segments, we define the probability of a particle landing on the $i$-th segment by the harmonic measure $\mu_i$, of the segment \cite{Schiffer}, as in DLA. The statistical weight that $k_i$ particles land to the $i$-th bin ($i = 1, 2, \dotsc, N$) within a time unit is given by the multinomial distribution:
\begin{equation}
    P(k_1, \dotsc, k_N) = N^K K! \prod_{i=1}^{N}\frac{\mu_i^{k_i}}{k_i!},     
\end{equation}
where $\sum_{i=1}^N \mu_i = 1$ and $\sum_{i=1}^N k_i = K$. The entropy functional defined as $S = \log P$ for the grown layer $\delta D$ in the Stirling approximation ($k_i \gg 1$) reads:
\begin{equation}\label{entropy-def}
    S[\delta D] = K\log N - \sum_{i=1}^N k_i\log(k_i/\mu_i K).   
\end{equation}

\textit{Equation of growth as the extremal of the entropy}. The entropy \eqref{entropy-def} is maximized when the particle count per segment is $k_i^* = K\mu_i$. To make a connection to the LG problem, we introduce a conformal map $z(t, w)$ from the exterior of the unit circle on the $w$-plane to the domain $D(t)$ on the physical $z$-plane with $z = x+iy$, such that the boundary of $D(t)$ is $z(t, \exp(i\phi))$ parametrized by $\phi \in [0, 2\pi]$. By equating a small displacement of the $i$-th bin of a boundary, $v_i \delta t$,  to $\sqrt\hbar\, k_i^* $, we immediately obtain the deterministic (classical) equation of growth
(see \cite{O1,O2} for details):

\begin{equation}\label{lg-equation}
    {\mathcal I}\text{m}\left(\bar z_t z_\phi\right) = Q/(2\pi),
\end{equation}
where subscripts denote partial derivatives. 

\textit{Entropy = scaled pattern area}.
The maximum entropy of the layer \eqref{entropy-def}, when $k_i^* = K\mu_i$, becomes $S[\delta D] = K\log N$, which is proportional to the area of the single layer, $K \hbar$. Because of additivity, the maximal entropy produced during $t/\delta t$ time steps corresponds (in fact, is proportional to) the domain's total area:
\begin{equation}\label{entropy}
    S[D(t)] = C\cdot Area[D(t)],
\end{equation}
where the domain independent constant $C = (\log N)/\hbar$. 

Since nature always favors the largest entropy scenario, then \eqref{entropy} implies that solving the selection problem is to find the value of the selecting parameter, which maximizes the area spanned by the growing domain $D(t)$.

\textit{Variational selection in a wedge}. A zero surface tension family of self-similar analytical solutions in sector (wedge) geometry with arbitrary opening angle $\theta$ has a form \cite{89,BA,Tu,AMZ-1}:
\begin{equation}\label{wedge-ss-solution}
    z(t, w) = r(t)w(1-w^{-2/\kappa})^\beta \pFq{2}{1}{\beta,\beta-\kappa}{1-\kappa}{w^{-2/\kappa}}.
\end{equation}
Here $z(t, \exp(i\phi))$ is an interface of a growing finger parameterized by the angle $\phi \in [0, 2 \pi)$, and $r(t) = \lim_{w\to\infty}|\partial_w z(t, w)|$ is a conformal radius of the conformal mapping. The solutions are labeled by the angle of the wedge, $\theta = \pi \kappa$, and by the angle, $\alpha = \pi (\kappa - \beta)$, between two tangents to the finger at the apex of the wedge.

The selection problem in wedge geometry is to find the observable (the most stable) member of the self-similar continuum \eqref{wedge-ss-solution}. The parameter of selection is the angle ratio
\begin{equation}\label{lambda}
    \lambda(\theta) = \alpha/\theta = 1 - \beta/\kappa,
\end{equation}
which becomes a function of $\theta$ after selection. Let us find $\lambda(\theta)$, which maximizes the area spanned by the growing domain $D(t)$. 

\begin{figure}[t]
    \centering
    \includegraphics[width=1\columnwidth]{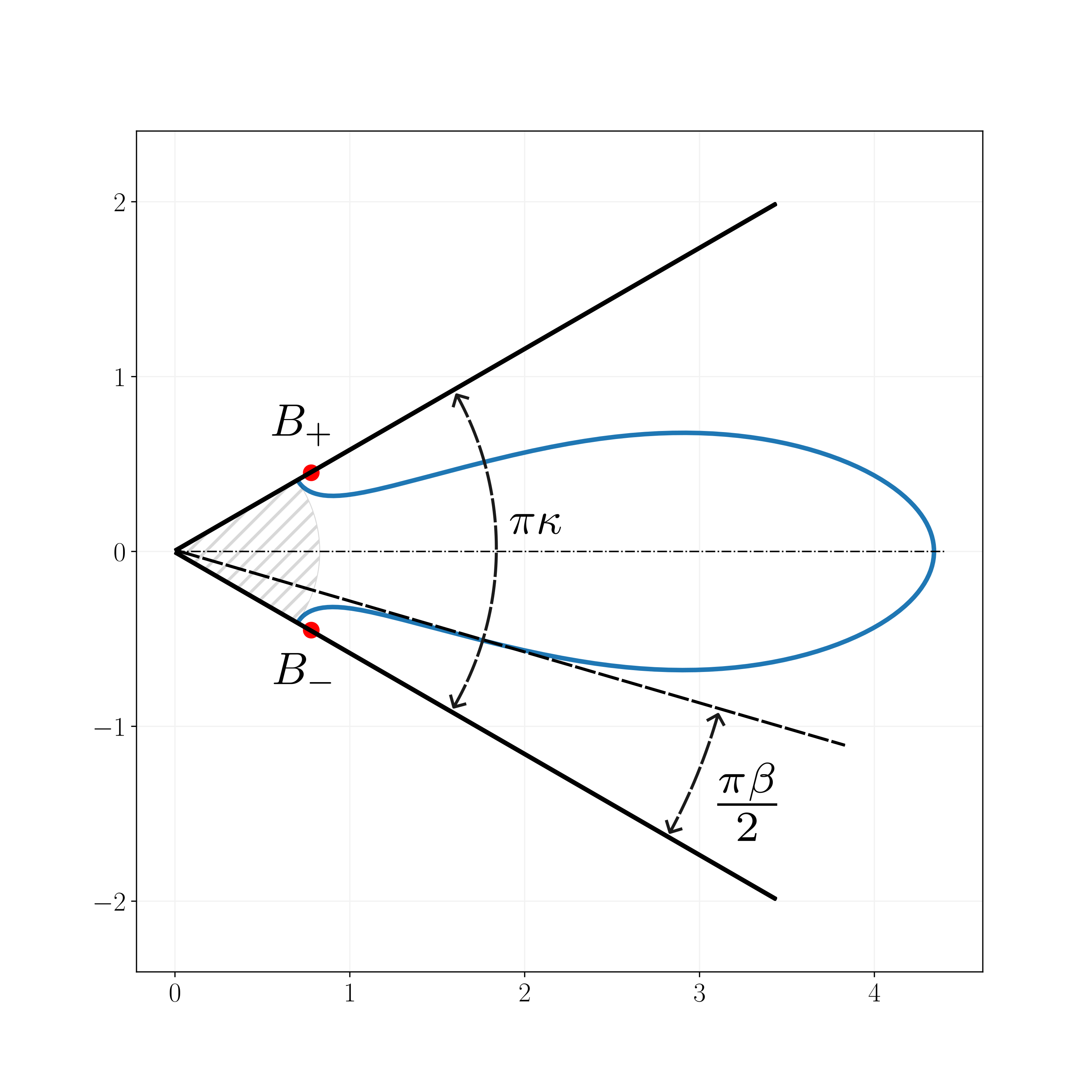}
    \caption{The finger in a wedge, described by eq. \eqref{wedge-t-solution}. The constants of motion, $B_{\pm} = B \exp(\pm i\pi \kappa /2)$, are shown by red circles. The dashed sector is the redundant part of the area.}
    \label{fig:wedge}
\end{figure}

\textit{Unsteady solution}. To apply the variational principle, one needs {\it unsteady} solutions, where the interface intersects the wedge walls instead of the unphysical self-intersection at the origin (a dashed line on \autoref{fig:wedge}), described by the self-similar family \eqref{wedge-ss-solution}.

By analyzing the evolution of isobars (level sets ahead of the self-similar interface), we derived an \textit{asymptotic} unsteady solution. This solution is valid when the shape of the evolving finger closely approximates the self-similar form given by \eqref{wedge-ss-solution}. The resulting solution is as follows:
\begin{equation}\label{wedge-t-solution}
    z = r\frac{w}{a}\left(1-\left(\frac{a}{w}\right)^{2/\kappa}\right)^\beta \pFq{2}{1}{\beta,\beta-\kappa}{1-\kappa}{\left(\frac{a}{w}\right)^{2/\kappa}},
\end{equation}
and the criterion of its validity is $1 - a \ll 1$\footnote{The solutions \eqref{wedge-ss-solution} and \eqref{wedge-t-solution} are valid until the first (inevitable) tip-splitting \cite{89,BA,Tu}.}. A typical interface described by this solution is schematically shown in blue in \autoref{fig:wedge}. 

The time evolution of $r(t)$ and $a(t)$ in \eqref{wedge-t-solution} can be found from the following two equations:
\begin{equation}
    \label{finger-Area}
    \mathcal A = \frac{1}{2i}\oint_{|w|=1} \bar z (1/w) d z(w) = Qt,
\end{equation}
where $\mathcal A$ is the finger area, and 
\begin{equation}
    \label{finger-B}
    B = r a^{-2}\left(1-a^{4/\kappa}\right)^\beta \pFq{2}{1}{\beta,\beta-\kappa}{1-\kappa}{a^{4/\kappa}},
\end{equation} 
where $B=z(t,1/a(t))$ is a time-independent constant, as follows from the integrability of LG and the Herglotz theorem on the singularity correspondence \cite{Herg,KMP,O1,O2}. The constant $B$ has the following interpretations:

(i) Geometrically, $B$ is a distance from the origin to the intersections of the interface with walls, minus a tiny correction, not written here to save space (see \autoref{fig:wedge}).

(ii) Analytically, $B$ is the branching point of a Riemann surface of a Schwarz function, defined at the interface as $\bar z = S(t, z) = \bar z (t, 1/w)$ (see details in \cite{Davis,OO}).

\begin{figure}[t]
    \centering
    \includegraphics[width=1\columnwidth]{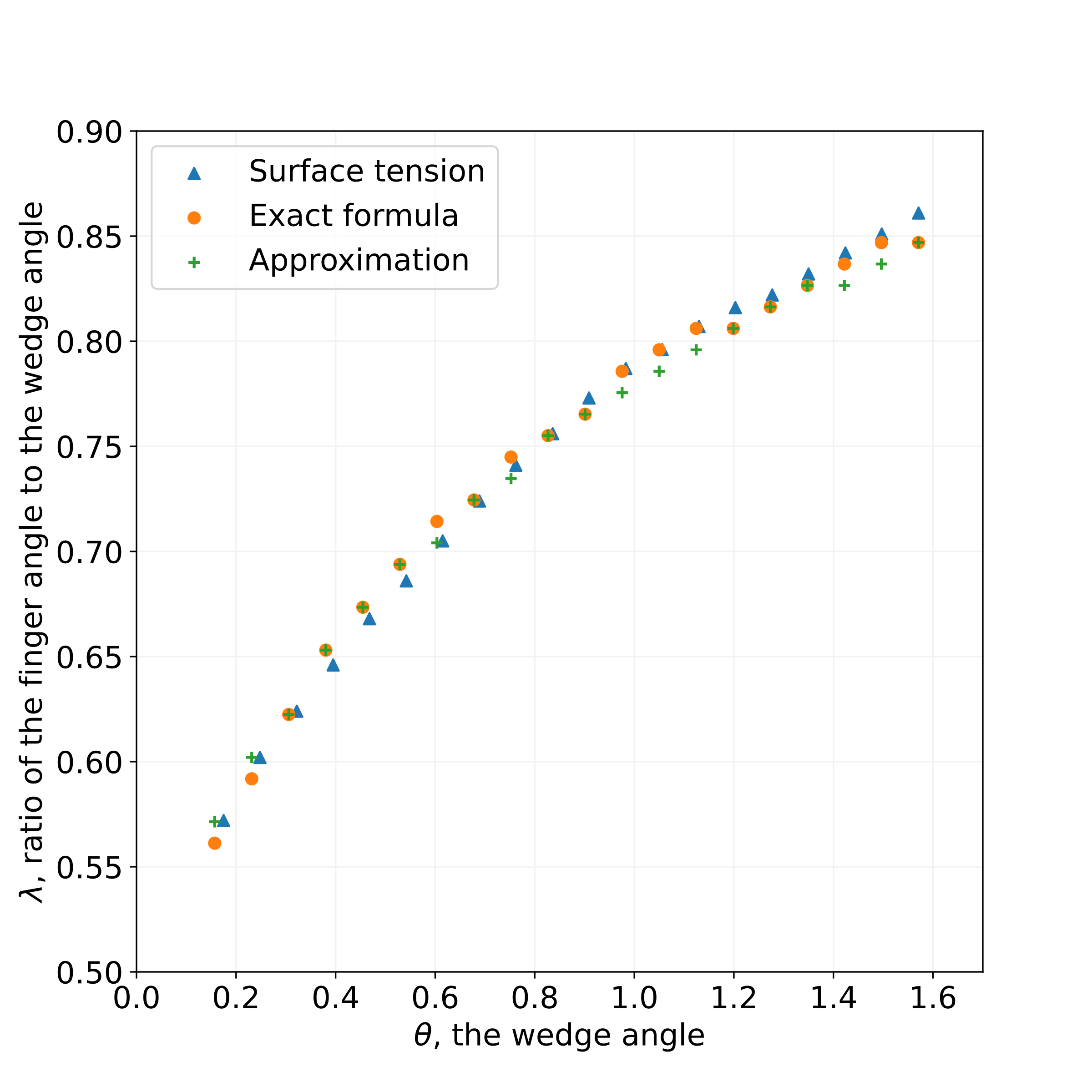}
    \caption{Blue triangles -- selection by surface tension \cite{Tu}, orange circles -- our selection by exact formula \eqref{wedge-area-selection}, green crosses -- selection by approximation \eqref{A0-appox}. The parameter $a(t)$ here equals $a(t)=1-10^{-7}$.}
    \label{fig:selection}
\end{figure}

\textit{Maximization of area}. To apply variational selection, the redundant part of the growing finger (the circular sector of the radius $B$ at \autoref{fig:wedge}) must be excluded. This region corresponds to a transient stage before the survival and formation of a single finger \eqref{wedge-t-solution}. The finger completes its formation and acquires the asymptotic shape when the intersection of the interface with the walls becomes exponentially close to the stagnation point $B$. Thus, the only area spanned by the asymptotic pattern and equaled the difference between the total area \eqref{finger-Area} and the sectorial area, $\pi B^2 \kappa / 2$, is considered for selection: 
\begin{multline}\label{wedge-area-selection}
    {\mathcal A}_0(\beta, \kappa, r(t), a(t))  = {\mathcal A}(\beta, \kappa, r(t), a(t)) - \\ - \pi B(\beta, \kappa, r(t), a(t))^2 \kappa / 2,
\end{multline}
We maximize $\mathcal A_0$ with respect to the fjord's opening angle $\beta$ at given $\kappa$, $r(t)$, and $a(t)$, and plot resulting $\lambda = \pi (1-\beta)/\theta$ as a function of the wedge angle $\theta$ in \autoref{fig:selection}. 

\textit{Results}. By comparing our results (orange circles) with results obtained by using the ``asymptotic beyond all orders'' with inclusion of surface tension \cite{Tu} (blue triangles) we observe an excellent agreement (the maximal discrepancy is about 1.5\%). 

The eq. \eqref{wedge-area-selection} is accurately approximated as
\begin{multline}\label{A0-appox}
    {\mathcal A}_0(\beta, \kappa, r, a) = \frac{\pi r^2\kappa}{2}\frac{\Gamma(1-\kappa)\Gamma(-\kappa)}{\Gamma(\beta-\kappa )\Gamma(1-\beta-\kappa)} - \\ -\frac{\pi r^2(\kappa-\beta)}{2(\kappa/4)^{2\beta}}\frac{\Gamma^2(1-\kappa)\Gamma^2(1-2\beta)}{\Gamma^2(1-\beta-\kappa)\Gamma^2(1-\beta)}(1-a)^{2\beta}.
\end{multline}
The error between \eqref{wedge-area-selection} and \eqref{A0-appox} is practically unnoticeable, as can be seen in \autoref{fig:selection}.

In a wedge with angle $\theta = \pi/2$, Eq. \eqref{A0-appox} can be simplified. In the leading order in $1-a\ll1$ it reads:
\begin{equation}
    {\mathcal A}_0(\beta, 1/2,r, a) =\frac{\pi r^2(1-2\beta)}{4}\left(\cos\pi\beta-2^{2\beta}(1-a)^{2\beta}\right).
\end{equation}
For $a = 1 - 10^{-7}$, this function reaches its maximum when $\beta = 0.077$, so that the selected $\lambda = 0.85$. It 
coincides with $\lambda$ in the article \cite{Brener90}, devoted to surface tension selection in the $90^\circ$ wedge, but differs by .01 from the value $\lambda = 0.86$ in \cite{Tu}, which is also selected by surface tension. 

\textit{Two unexpected observations}. When $a(t)$ is not close to 1 the finger has not yet formed its self-similar asymptotic shape \eqref{wedge-ss-solution}, so it is too early to test a pattern for selection: one should wait until growing $a(t)$ enters the vicinity of 1, so that $1 - a \ll 1$. Then the shape of the finger is finally formed at $a(t)=1-10^{-7}$ (as shown in \autoref{fig:selection}), and subsequent evolution is not expected to change the selected value of $\beta$ at given $\kappa$. For earlier times, the shape has not yet formed, as said above, so it gives less than an asymptotic value of $\beta$.  

(i) But, as we unexpectedly found, for higher values of $a(t)$, that is, for higher times, the selected value of $\beta$ gradually loses its precision. This is because the area ${\mathcal A}$ grows in time contrary to the second term in the RHS of \eqref{wedge-area-selection}, which stays constant in time, so in a long time limit a contribution of the second term in \eqref{wedge-area-selection}, which is crucial for selection, becomes negligible. This explains why, we believe, the accuracy of the selected value worsens for very large $t$, and at the limit $t \to \infty$ the selection disappears completely. 

(ii) It is also surprising that the critical value $a(t)=1-10^{-7}$ is the same for all wedge angles, $0^\circ < \theta \leq 90^\circ$. 

Future work on selection in a wedge must shed light on both of these unexpected observations.

\textit{No attractor}. Motivated by the success of selections in a channel \cite{98,TS-1,Robb,TS-2} without surface tension, we expected the attractor of \eqref{lg-equation} to be the selected pattern in a wedge. Since we did not find the attractor (maybe there is no such in a wedge, contrary to a channel), additional information beyond Eq. \eqref{lg-equation} was needed for selection. This additional information appears to be the entropy \eqref{entropy} in a form of the pattern area, ${\mathcal A}_0$, in \eqref{wedge-area-selection} associated with each member of the continuous family \eqref{wedge-t-solution}.

\begin{figure}[t]
    \centering
    \includegraphics[width=1\columnwidth]{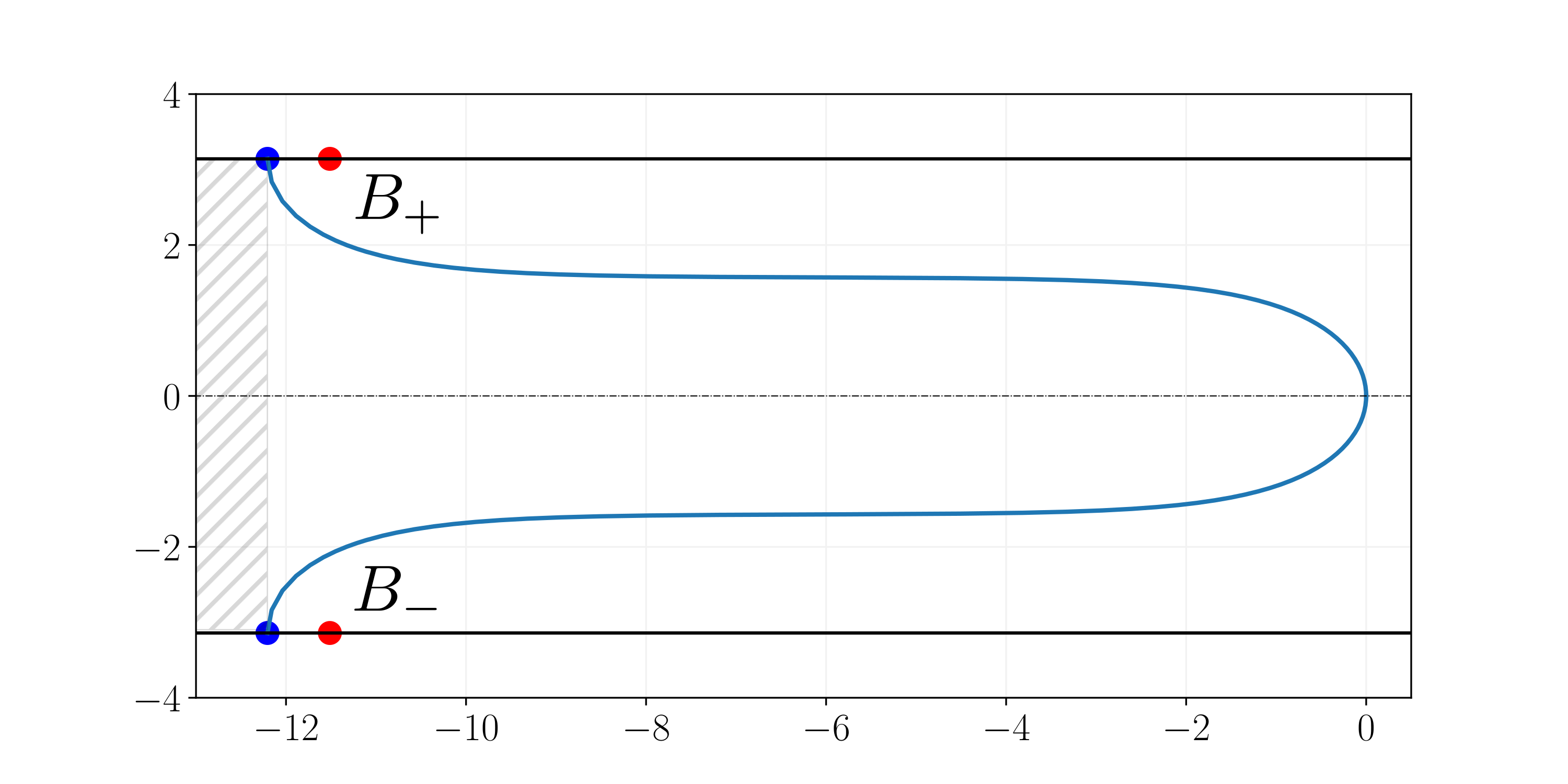}
    \caption{Moving finger in a rectangular channel, described by eq. \eqref{finger-ss-solution}. The constants of motion, $B_{\pm} = B \pm i \pi $, are indicated by red circles. The stagnation points, $B_{\pm} - 2(1 - \lambda) \log 2$, are shown by blue circles. The dashed rectangle is the redundant part of the total finger area.}
    \label{fig:finger}
\end{figure}

\textit{Finger selection in a channel}. For completeness, we briefly address the STF selection problem in the channel geometry $\theta \to 0$. We cannot apply the maximal area principle to the family of the Saffman-Taylor fingers
\begin{equation}\label{finger-ss-solution}
x = Ut + 2(1 - \lambda) \log(\cos(y/2\lambda)),    
\end{equation}
since they are infinitely long, so their interior area diverges. Fortunately, there is an unsteady solution for a finite finger \cite{S59,ms,sm,kup,98} shown in \autoref{fig:finger}:
\begin{equation}\label{finger-t-solution}
    z = r(t) + i\phi + 2(1 -  \lambda) \log (1 - a(t) \exp(-i\phi)),    
\end{equation}
where $r(t)$ and $a(t)$ can be found from the equations:
\begin{gather}
    \label{channel-area}
    {\mathcal A} = t = r(t) + 2(1 - \lambda)^2 \log(1 - a(t)^2),\\
    \label{channel-B}
    B = z(t, 1/a(t)) = r(t) + 2(1  - \lambda)\log(1 - a(t)^2).
\end{gather}
Here ${\mathcal A}$ is the area inside the finger, which equals time, $t$, in our scaled units, $Q = 2\pi$, in \eqref{lg-equation}, $B$ is a constant of motion \cite{Herg,KMP,O1,O2}, $r(t)$ is the moving fingertip, and $a(t)$ is related to the length of the finger, which diverges when $a(t) \to 1$. Initially, at $t = 0$, the finger is almost flat, so $a(0) \ll 1$. Then $a$ starts to grow and because $da(t)/dt > 0$ moves (exponentially slow when $t \to \infty$) toward 1.

After eliminating $r(t)$ in these two equations by subtracting $B$ from ${\mathcal A}$ in \eqref{channel-area}, we obtain:
\begin{equation}\label{finger-area}
    {\mathcal A}_0(\lambda, a(t)) =  -2 \lambda (1 - \lambda) \log (1 - a(t)^2).
\end{equation}
Geometrically, this subtraction is a removal of the redundant part of the total area of the finger, that is, the rectangle, shown in \autoref{fig:finger}.

As Eq. \eqref{finger-area} shows, the area, ${\mathcal A}(\lambda, a(t))$, is maximal at $\lambda = 1/2$, which is the selected value according to the classical experiment \cite{ST58}. This selection is achieved by a simple variational principle, which favors the most probable scenario (see above) and amounts to the maximal area spanned by the evolving pattern.

\begin{figure}[t]
    \centering
    \includegraphics[width=1\columnwidth]{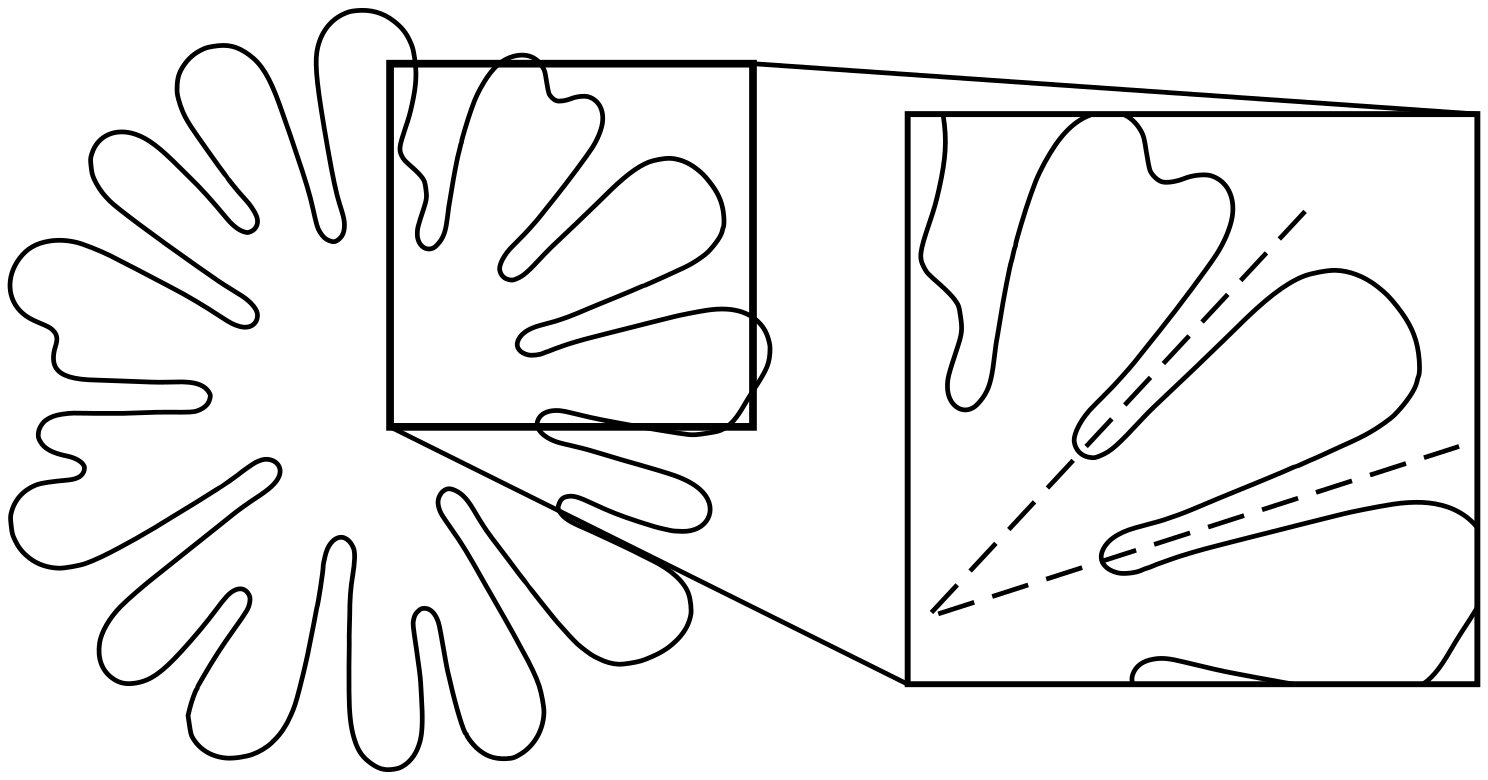}
    \caption{A wedge-like fragment of the interface in radial geometry (between dashed lines). The centerlines of two fjords surrounding the finger form the walls of this virtual wedge.}
    \label{fig:wedge-zoom}
\end{figure}

\textit{Universal fjord opening angle.}
Researchers at the University of Texas shifted the focus from viscous fingers to fjords, which separate growing fingers \cite{2OO6}. These fjords are stable and robust, in contrast to less predictable fingers due to their unstable evolution. Another (and not less notable) feature of the fjords is that they are the building blocks of exact solutions of the Laplacian growth equation without surface tension (see \autoref{fig:wedge-zoom}). More precisely, all the geometric characteristics of the fjords (vertices, locations, directions, and shape details) are constants of motion of eq. \eqref{lg-equation} represented by the singularities of the Schwarz function \cite{ms, sm, AMZ-1, AMZ-2, MPT}.

A key discovery was the universal opening angle of the fjord, measured experimentally as $8.0^\circ \pm 1.0^\circ$, consistent with various pumping rates and fjord lengths, widths and directions \cite{2OO6}. If the fjord's opening angle is universal, $\pi \beta_0$, for a noticeable range of the wedge angle $\theta$, then the plot in \autoref{fig:selection} should fit the formula, 
\begin{equation}\label{lambda-opening}
    \lambda(\theta) = 1 - \pi\beta_0/\theta,    
\end{equation}

This law was also experimentally discovered in the wedge geometry \cite{89}: $\lambda(\theta) = 1 - 10^\circ / \theta$. The authors of \cite{89} write about this formula, obtained from their measurements: ``This empirical law is unexplained and could be due to a mere coincidence.'' 

\textit{Our results}. The formula \eqref{lambda-opening} indeed fits our plot in \autoref{fig:selection} well for $\pi\beta_0 = 11.7^\circ$ in the range $35^\circ<\theta<90^\circ$, and coincides with one obtained numerically by using surface tension in \cite{Tu}. However, our result cannot explain the discrepancy with somewhat lower experimental values $10.0^\circ$ for real \cite{89} and $8.0^\circ \pm 1.0^\circ$ and for virtual \cite{2OO6} wedges, respectively.
 
\textit{Conclusion.} A straightforward variational principle was derived and applied to pattern selection problems in a Hele-Shaw cell in wedge and channel geometries. The results obtained are in excellent agreement with the experiments. This principle complements a non-variational selection developed earlier \cite{98}, since for both selections, surface tension is not needed, contrary to common belief.

The authors acknowledge useful discussions with J. Pearson and V. Fradkov. The work of O.A. was supported by the Russian Science Foundation grant 19-71-30002.

\bibliographystyle{ieeetr}
\bibliography{biblio}{}

\end{document}